\documentclass[prl,twocolumn,superscriptaddress,showpacs,floatfix]{revtex4-2} 

\usepackage{hyperref}
\usepackage{dcolumn} 
\usepackage{bm} 
\usepackage{amssymb} 
\usepackage{amsmath}
\usepackage{color} 
\usepackage{graphics}

\usepackage{epsfig}
\usepackage[version=4]{mhchem}
\begin{document}
\title{The Influence on Crystal Nucleation of an Order-Disorder Transition among the Subcritical Clusters.}

\author{Richard K. Bowles}
\email{richard.bowles@usask.ca}
\affiliation{Department of Chemistry, University of Saskatchewan, Saskatoon, SK, S7H 0H1, Canada.}
\affiliation{Centre for Quantum Topology and its Applications (quanTA), University of Saskatchewan, SK S7N 5E6, Canada.}

\author{Peter Harrowell}
\email{peter.harrowell@sydney.edu.au}
\affiliation{School of Chemistry, University of Sydney, New South Wales 2006, Australia}

\date{\today}

\begin{abstract}
Studies of nucleation generally focus on the properties of the critical cluster, but the presence of defects within the crystal lattice means that the population of nuclei necessarily evolve through a distribution of pre-critical clusters with varying degrees of structural disorder on their way to forming a growing stable crystal. To investigate the role pre-critical clusters play in nucleation, we develop a simple thermodynamic model for crystal nucleation in terms of cluster size and the degree of cluster order that allows us to alter the work of forming the pre-critical clusters without effecting the properties of the critical cluster. The steady state and transient nucleation behaviour of the system are then studied numerically, for different microscopic ordering kinetics. We find that the models exhibits a generic order-disorder transition in the pre-critical clusters. Independent of the type of ordering kinetics, increasing the accessibility of disordered pre-critical clusters decreases both the steady state nucleation rate and the nucleation lag time. Furthermore, the interplay between the free energy surface and the microscopic ordering kinetics leads to three distinct nucleation pathways.
\end{abstract}
\maketitle

There is growing evidence to suggest that many systems crystallize via non-classical nucleation where fluctuations other than the critical nuclei play a significant mechanistic role in the phase change~\cite{Wolde_Enhancement_1997,Talanquer_Crystal_1998,Whitelam_Nonclassical_2010,Harrowell_2010,Lutsko_How_2019}. For example, two-step nucleation~\cite{Duff_Nucleation_2009,Iwamatsu_Free_2011,Lester_Patterning_2012,Malek_Crystallization_2015,James_Phase_2019,Poole_Free_2021} , which has been observed variety of molecular~\cite{Gebauer_Stable_2008,Sear_The_2013,Ishizuka_Two_Step_2016,Kumar_Two-Step_2018}, colloidal systems~\cite{Savage_Experimental_2009,Peng_Visualizing_2014,Haxton_Crystallization_2015,Peng_Two-Step_2015,Qi_Nonclassical_2015,Lee_Entropic_2019}, involves the initial formation of small pre-critical clusters structurally related to thermodynamically nearby metastable states that then transform to the stable state at a larger size.  The steady state value of the nucleation rate is typically regarded as being entirely determined by the kinetics at the critical nucleus~\cite{Kelton_Book_2010}. This would appear to leave only the transient nucleation rate  (i.e. on the approach to steady state) to reflect the ordering process within pre-critical clusters. In this paper, we show that the accessibility of disordered pre-critical clusters affects both the steady state crystal nucleation rate and transient nucleation behaviour, while competition between the microscopic ordering kinetic and the geometry of the free energy surface can lead to different nucleation mechanisms.

Disorder in a crystal cluster will have two generic consequences. One will be to increase the bulk contribution to free energy of the cluster over that of the perfect crystal. The second consequence will be to decrease the crystal-liquid interfacial free energy by diminishing the entropic difference between the adjacent phases. Here, we introduce a simple model that captures these two competing effects and describes the reduced work of forming a cluster as,
\begin{equation}
\begin{split}
\Delta f(&n,\phi)=\frac{\Delta G(n,\phi)}{|\Delta\mu_c| k_B T}\\
&=\Delta_d (1-\phi)n-\phi n+\gamma_c\sigma(1-(1-\phi)\delta)n^{2/3}\mbox{,}
\label{eq:fe}
\end{split}
\end{equation}
where $\Delta G$ is the Gibbs free energy of forming a cluster of size $n$, with a degree of order, $\phi$, and $\Delta\mu_c$ is the difference in chemical potential between the stable equilibrium crystal phase and the metastable liquid. When $\phi=1.0$, the cluster has the structure of the perfect crystal, but $\phi$ is decreased by the presence of defects characterized by bulk excess free energy relative to that of the equilibrium crystal, $\Delta_d>0$. While the structural nature of the defect can be interpreted broadly, it is necessarily distinct from the average order of the metastable fluid so that a cluster with $\phi=0$ is structurally distinct from the fluid and thermodynamically unstable. The surface contributions to $\Delta f$ are given by the reduced surface free energy of the perfectly ordered crystal, $\gamma_c$, the relative surface free energy decrease due to disorder, $\delta$ and a geometric factor, $\sigma$ that accounts for the shape of the cluster.

\begin{figure*}[t]
\includegraphics[]{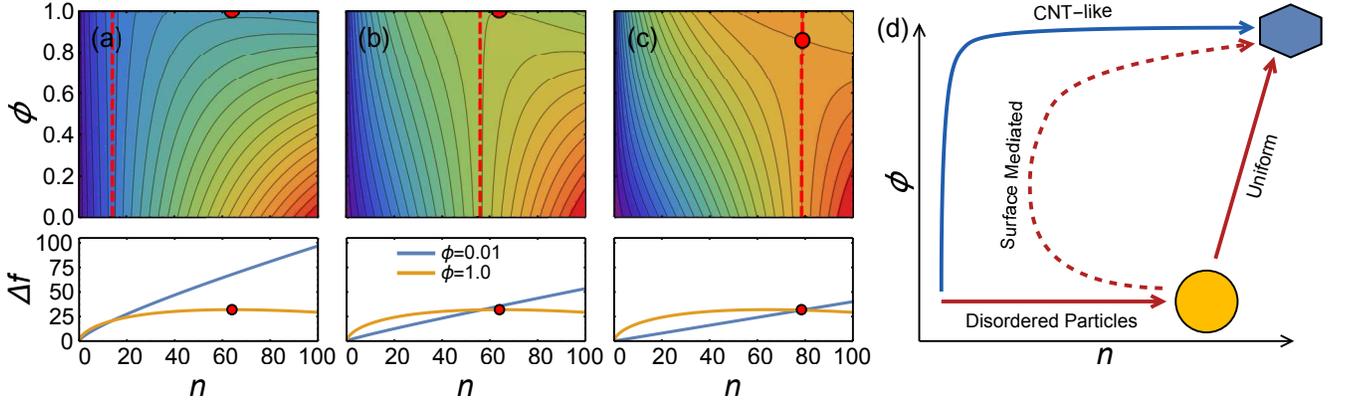}
\caption{Free energy, $\Delta f(n,\phi)$ (Eq.~\ref{eq:fe}), using $\Delta_d=0.4$ and $\gamma_c\sigma=6.0$, for (a) $g=0.40$, (b) $g=0.64$ and (c) $g=0.71$.  (Top) Contour plots of the free energy as a function $n$ and $\phi$, with $n^{\dagger}$ and $n^*$ denoted by a red dashed line and red point, respectively.  (Bottom) Free energy profile as a function of $n$ for disordered clusters, $\phi=0.01$, and ordered clusters, $\phi=1.0$. (d) Nucleation pathways through the order parameter space of the free energy surface. See text for details. } 
\label{fig:fesurface}
\end{figure*}

The description of the free energy can simplified by introducing the parameter  $g=\delta/(1-\Delta_d)$ and Figs.~\ref{fig:fesurface}(a)-(c) show the free energy surface for different values $g$, along with the free energy profiles for the growth of ordered and disordered clusters. A key feature of the model is the appearance of an order-disorder transition in the pre-critical clusters. With $\phi=1.0$, Eq.~\ref{eq:fe} reduces usual expression for Classical Nucleation Theory (CNT), $\Delta G(n)=-\Delta\mu_c n+\gamma_c\sigma n^{2/3}$, going through a maximum at the critical size $n^*=8(\gamma_c\sigma)^3/27$, and is independent of $g$. At small $\phi$, the free energy increases monotonically with $n$, indicating disordered clusters are always thermodynamically unstable with respect to cluster size and will tend to shrink. The free energy is linear in $\phi$ at fixed $n$.  When a cluster is small, the slope at fixed $n$, $(\partial \Delta f/\partial \phi)_n$, is positive because the surface free energy cost of forming an ordered solid-fluid interface is significantly greater than that for the disordered clusters, while the respective gain in bulk free energy is small. For larger clusters, the greater bulk free energy gain of the ordered state begins to dominates and we see a cross over at a cluster size, $n^{\dagger}=(\gamma_c\sigma)^3g^3=\frac{27}{8}n^*g^3$, where the slope becomes negative and the ordered clusters become more stable.

For $g\le2/3$, $n^{\dagger}$ occurs before $n^*$, which is independent of $g$, and the lowest free energy path for nucleation involves the initial growth of small disordered clusters that then order at $n^{\dagger}$ to form a perfectly ordered clusters before eventually going over the nucleation barrier (Figs.~\ref{fig:fesurface}(a) and (b)). Phenomenologically, the model is similar to two-step nucleation models where the intermediate metastable phase is higher in free energy than the mother phase and has low surface tension~\cite{Iwamatsu_Free_2011,Poole_Free_2021}, but the defects included here cannot form a bulk thermodynamic phase so the clusters remain thermodynamically unstable and the key features of the model should be generally applicable to crystal nucleation.
For $g\ge 2/3$, $n^*$ begins to grow and become more disordered and  $n^{\dagger}=n^*$ (Figs.~\ref{fig:fesurface}(c)). We  focus on cases with $n^{\dagger}<n^*$.

The kinetic evolution of clusters on this free energy surface can be characterized in terms equilibrium reactions for the growth and decay of a cluster through the addition or loss of a monomer and for the order--disorder processes. The rate constants for the growth and decay processes, obtained from a simple rate theory, are given by~\cite{Turnbull_Rate_1949,Kelton_Transient_1983,Kelton_Book_2010},
\begin{align}
\kappa^+_{n,\phi}&= n^{2/3}\exp\left[-\Delta_nG(n,\phi)/2k_BT\right]\mbox{,}\nonumber\\
\kappa^-_{n,\phi}&= (n-1)^{2/3}\exp\left[\Delta_nG(n-1,\phi)/2k_BT\right]\mbox{,}
\label{eq:knp}
\end{align}
where  $\Delta_nG(n,\phi)=\Delta G(n+1,\phi)-\Delta G(n,\phi)$ and the prefactor accounts for the addition and loss of monomers at the surface. Similarly, the rate constants for the order-disorder kinetics are, 
\begin{align}
\omega^+_{n,\phi}&= \alpha(n)\exp\left[-\Delta_{\phi}G(n,\phi)/2k_BT\right]\mbox{,}\nonumber\\
\omega^-_{n,\phi}&= \alpha(n)\exp\left[\Delta_{\phi}G(n,\phi-\Delta\phi)/2k_BT\right]\mbox{,}
\label{eq:wnp}
\end{align}
where $\Delta_{\phi}G(n,\phi)=\Delta G(n,\phi+\Delta\phi)-\Delta G(n,\phi)$, and $\alpha(n)$ is a size dependent prefactor that captures the effects of different ordering mechanisms. The ordering kinetics in a cluster will depend both on the relationship between the equilibrium crystal order and the local structure of the defect, and where structural relaxation can occur. The crystal-fluid interface at the cluster surface is the most dynamic region of a cluster, where annealing can occur. If the defects structure is incompatible with the crystal structure so the defect must diffuse to the surface before rearranging in a surface mediated ordering (SMO) process, then $\alpha(n)\sim D/n^{2/3}$, where the size dependence accounts for the additional time it takes for a defect to reach the surface in larger clusters and $D$ is the defect diffusion coefficient. As an alternative, we assume a 
uniform ordering (UO) process, were the defects can easily rearrange without diffusion, has kinetics determined by the free energy surface alone and $\alpha(n)=C$ is size independent constant.


The time dependent forward rates of cluster growth and cluster ordering can be defined for $n,\phi$--clusters as,
\begin{align}
I^{g}_{n,\phi,t}&=\kappa^+_{n,\phi}N_{n,\phi,t}-\kappa^-_{n+1,\phi}N_{n+1,\phi,t}\mbox{,}\nonumber\\
I^{o}_{n,\phi,t}&=\omega^+_{n,\phi}N_{n,\phi,t}-\omega^-_{n,\phi+\delta\phi}N_{n,\phi+\delta\phi,t}\mbox{,}
\label{eq:fn}
\end{align}
respectively, where $N_{n,\phi,t}$ is the number of clusters with size and order $n,\phi$ at time $t$. The time evolution of the cluster population can then be described by,
\begin{equation}
\frac{\partial N_{n,\phi,t}}{\partial t}=I^g_{n-1,\phi,t}-I^g_{n,\phi,t}+I^o_{n,\phi-\delta\phi,t}-I^o_{n,\phi,t}\mbox{,}\\
\label{eq:dNnt}
\end{equation}
which we solve numerically using discrete time intervals, $\delta t$, following a method developed by Kelton et al~\cite{Kelton_Transient_1983}. Full details of the numerical method can be found in the Supplementary Information (SI)~\cite{SI}.

\begin{figure}[t]
\includegraphics[]{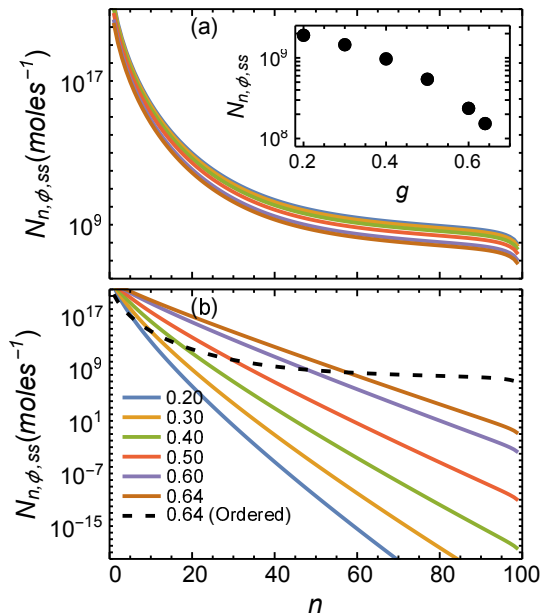}
\caption{Steady state cluster populations, $N_{n,\phi,ss}$ as a function of $n$ in the surface mediated ordering model with different values of $g$ for (a) ordered clusters $\phi=1.0$ (Inset: $N_{n,\phi,ss}$ for critical clusters, $n^*=64$ and $\phi=1.0$ as a function of $g$) and (b) disordered clusters, $\phi=0.01$. The dash line represents the ordered cluster for comparison.}
\label{fig:NNSSc}
\end{figure}

The steady state population of clusters, $N_{n,\phi,ss}$, obtained at $t\approx3000$s, is solely a function of the free energy surface and is the same for both kinetic models. Figure~\ref{fig:NNSSc}a shows that $N_{n,\phi,ss}$ for the ordered clusters exhibits a decay consistent with the expectations for the steady state population from a CNT free energy model. At $\phi=1.0$, $\Delta f(n,\phi)$ is independent of $g$, so the small decrease in the steady state population at the transition state as $g$ increases (Inset: Fig.~\ref{fig:NNSSc}a) is the result of changes in the free energy at other points on the surface. However, increasing $g$ directly lowers  $\Delta f(n,\phi)$ for the disordered clusters, making these states more accessible, and causing $N_{n,\phi,ss}$ to increases by many orders of magnitude. Fig.~\ref{fig:NNSSc}b shows the large excess of disordered clusters relative to ordered clusters for intermediate cluster sizes. 

\begin{figure}[t]
\includegraphics[]{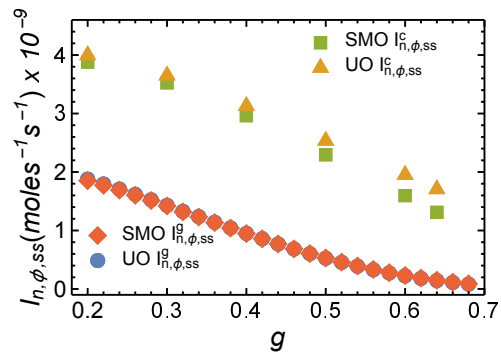}
\caption{$I^g_{n^*,\phi^*,ss}$ and  $I^c_{n,\phi,ss}$ as a function of $g$ for SMO and UO microscopic ordering kinetics.}
\label{fig:ISS}
\end{figure}

To examine the effect this increased population of disordered clusters has on the nucleation rate, we measure the steady state forward growth rate through the critical cluster, $I^g_{n^*,\phi^*,ss}$, and because the two dimensional nature of free energy surface means clusters can nucleate without passing through the critical cluster, we also measure the total steady state forward rate entering the free energy crystal basin, $I^c_{n,\phi,ss}$, where, for simplicity, the crystal basin is defined as the rectangular region $n>n^*, \phi > 0.9$. Figure~\ref{fig:ISS} shows that $I^g_{n^*,\phi^*,ss}$ and  $I^c_{n,\phi,ss}$ decrease by factors of approximately 10 and 2, respectively, as $g$ increases. The effect is independent of the nature of the microscopic kinetics, although there is a small but growing difference in $I^c_{n,\phi,ss}$ for the SMO and UO kinetics at large $g$, which suggests it arises from the changes in the free energy surface and the accompanying increase in the accessibility of the disordered cluster states. One way to understand the rate decrease is to note that the pre-critical clusters represent states associated with the metastable liquid. By increasing the configuration space associated with the metastable state, the probability of the system being at the critical state is decreased, effectively increasing the nucleation barrier~\cite{Scheifele_Heterogeneous_2013,Asuquo_Competitive_2016}.

\begin{figure}[t]
\includegraphics[]{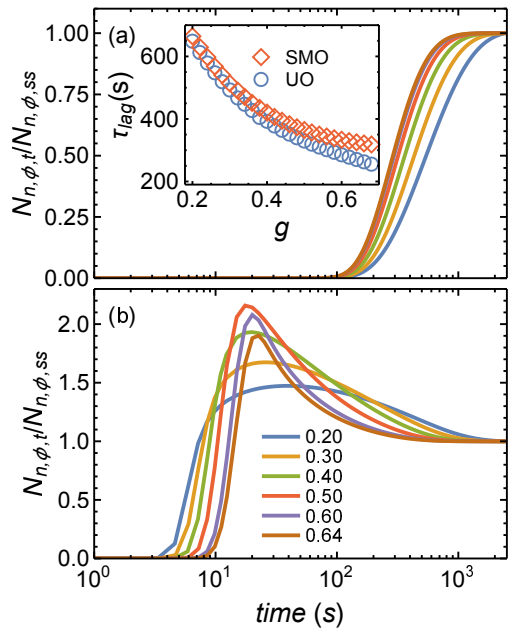}
\caption{Relative transient cluster populations for $n=64$ sized clusters as a function of time in the SMO model with different values of $g$ (Similar results for the UO model not shown). (a) ordered clusters $\phi=1.0$. Inset shows $\tau_{lag}$ as a function of $g$ for uniform and surface--mediated ordering. (b) Disordered clusters $\phi=0.01$.}
\label{fig:Nntrans}
\end{figure}

\begin{figure*}[t]
\includegraphics[width=5.5in]{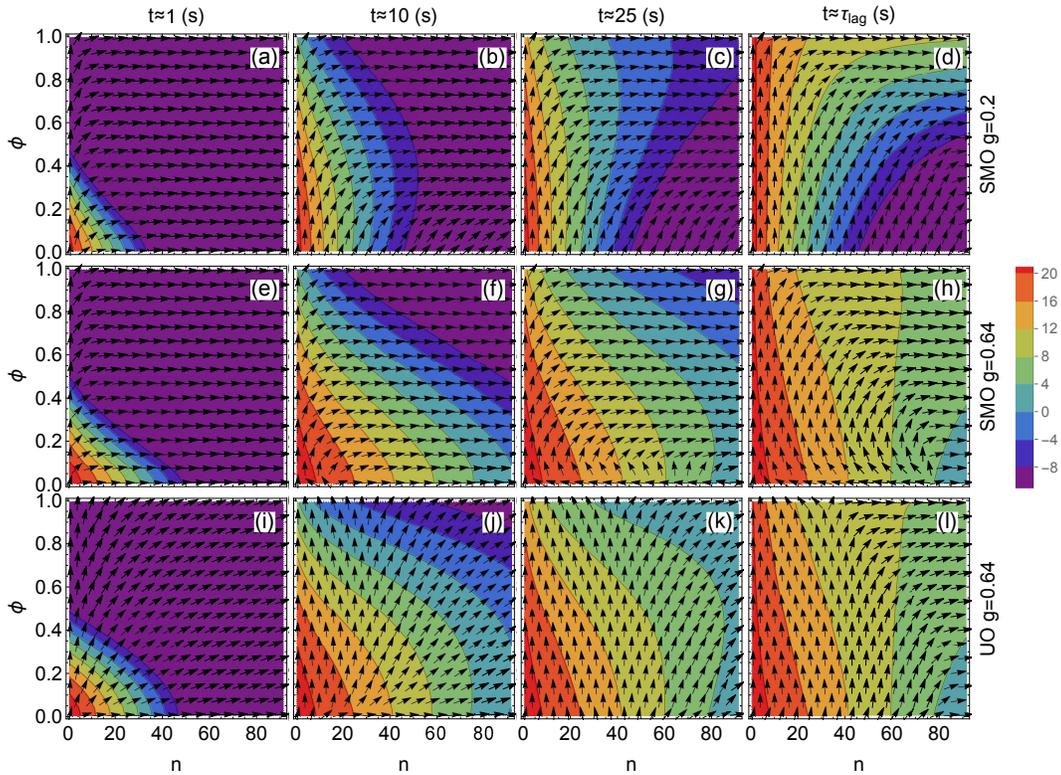}
\caption{Contour plots showing $\ln N_{n,\phi,t}$ and vector plots of the forward rates for (a-d) SMO with $g=0.2$, (e-h) SMO with $g=0.64$ and (i-l) UO with $g=0.64$, at times $t\approx 1\mbox{ s}, 10\mbox{ s}, 25\mbox{ s}$ and $\tau_{lag}$ for columbs left to right, respectively.  The vector components, $I^{g}_{n,\phi,t}$ and $I^{o}_{n,\phi,t}$, are normalized relative to $|I_{n,\phi,t}|=\sqrt{(I^{g}_{n,\phi,t})^2+(I^{o}_{n,\phi,t})^2}$.}
\label{fig:vectors}
\end{figure*}

Increasing the accessibility of the disordered cluster states also influences the transient nucleation behaviour, with Fig.~\ref{fig:Nntrans}(a) showing that the transient cluster population, relative to steady state value, for the critical cluster grows at earlier times larger $g$, leading to lower lag times, $\tau_{lag}$ (see Inset). To understand this, we follow the transient cluster populations of the disordered clusters as a function time. Figure~\ref{fig:Nntrans}(b) shows that $N_{n,\phi,t}$ for large disordered clusters in the SMO model actually overshoot their steady state populations at times an order of magnitude earlier than the critical clusters, then decay towards the steady state at longer time scales. However, for $g=0.2$ these large disordered populations are very small and the cluster population is restricted to a narrow channel for small disorder clusters, so clusters must order before they grow (Fig.~\ref{fig:vectors}(a-d)). With $g=0.64$, large excess populations of disordered clusters form at early times that then flow on to other regions of the free energy surface through an increased number accessible cluster states (Fig.~\ref{fig:vectors}(e-h)) to decrease $\tau_{lag}$.

Figure~\ref{fig:vectors} also highlights the interplay between the free energy surface and the microscopic kinetics in determining the nature fo the nucleation mechanism, which in the present system yields the three nucleation paths ways depicted in Fig.~\ref{fig:fesurface}(d). With $g=0.2$, both the SMO ((Fig.~\ref{fig:vectors}(a-d)) and UO models exhibit the same CNT-like nucleation mechanism because both the free energy surface and the microscopic kinetics of the SMO model favour ordering at small clusters. When $g=0.64$, the free energy surface favours formation of large disordered clusters, and we see large populations of disordered clusters form at early times, but the subsequent evolution of the clusters is determined by the microscopic kinetics. The ordering kinetics is suppressed for larger disordered clusters in the SMO model because defects must diffuse to the surface before they can rearrange to the stable state. As a result, the growth rate, $I^g_{n,\phi,t}$ for these disordered clusters becomes negative as the cluster tend to shrink to smaller sizes before ordering, as indicated by the rate vector field (Fig.~\ref{fig:vectors}(e-h)) (See SI for more details). In contrast, the ordering process for the large disordered clusters in the UO model is determined by the free energy surface, which is downhill for $n>n^{\dagger}$, and the large disordered cluster tend to order directly towards crystal. The more direct route over the free energy surface for the disordered clusters in the UO model may also account for the model's shorter lag times compare to those in the SMO model.

In conclusion, we have shown that the inclusion of additional order parameter(s) describing the clusters coupled with a free energy that promotes the sampling of this expanded configuration space will produce, quite generally, a reduction in the steady state nucleation rate by reducing the amount of available monomer. Here, we only included one defect type. Increasing the number of different types of defects and disorder available to the crystal would expand the number of available pre-critical states, which could lead to further slowing of the nucleation rate that has implications for the glass forming ability of a material. Beyond this generic impact, we have shown how the details of the ordering kinetics in the sub-critical clusters can significantly alter the transient nucleation kinetics. This is of particular importance in the analysis of molecular dynamics simulations of crystal nucleation where the transients will dominate over the accessible observation time. Furthermore, while large disordered clusters has been observed, for example in protein crystallization~\cite{Veklov_The_2010,Zhang_Nonclassical_2017,Houben_A_2020}, its not clear if they are directly involved in the  nucleation pathway. Our work suggests that the nature of the microscopic growth and ordering kinetics might provide insight into the role these these large disordered clusters play in nucleation.

\begin{acknowledgments} 
We would like to acknowledge NSERC grant RGPIN--2019--03970 (R. K. B) for financial support. Computational resources were provided by WestGrid and Compute Canada.
\end{acknowledgments}

%




\clearpage
\widetext
\begin{center}
\textbf{\large Supplemental Material: The Influence on Crystal Nucleation of an Order-Disorder Transition among the Subcritical Clusters.}
\end{center}
%
\setcounter{equation}{0}
\setcounter{figure}{0}
\setcounter{table}{0}
\setcounter{page}{1}
\makeatletter
\renewcommand{\theequation}{S\arabic{equation}}
\renewcommand{\thefigure}{S\arabic{figure}}
Here, we provide supplemental material  in support of our results, including additional details of the numerical methods used to calculate the transient and steady state nucleation rates for the model, and some additional analysis of the cluster populations.

\section{Numerical Simulation Details}
Equation~\ref{eq:dNnt} is solved numerically using discrete time steps, $\delta t$, so the population at time $t+\delta t$ is given by,
\begin{equation}
\begin{split}
 N_{n,\phi,t+\delta t}&=N_{n,\phi,t}+\\
 &\delta t\left[I^g_{n-1,\phi,t}-I^g_{n,\phi,t}+I^o_{n,\phi-\delta\phi,t}-I^o_{n,\phi,t}\right]\mbox{.}
 \end{split}
  \label{eq:Nnt}
 \end{equation}
Equation~\ref{eq:Nnt} is only exact in the limit $\delta t\rightarrow 0$, but provides a reasonable approximation to the solution of Eq.~\ref{eq:dNnt} as long as $\delta t$ remains small. Following the method developed by Kelton et al~\cite{Kelton_Transient_1983},  we begin with $\delta t=5.0\times 10^{-9}$s. The step size is then increased by $3.0\delta t$, up to a maximum time step $5.0\times 10^{-5}$s, when the new $\delta t$ is able to accurately predict the total cluster population over the last three successive time steps within $0.1\%$. Our simulations are run until $t=3000$s.

The the reduced simulation time $t^*=t\gamma$ is controlled by the unbiased molecular jump rate in the liquid,
\begin{equation}
\gamma=\frac{6D_l}{\lambda^2}\mbox{,}\\
\label{eq:ljump}
\end{equation}
where $D_l$ is the diffusion coefficient in the liquid and $\lambda$ is the molecular jump distance. For convenience, we simply take $\gamma=1$s$^{-1}$, which, to give this value some context, would describe the dynamics in liquid lithium disilicate  at $T\approx800$K~\cite{Kelton_Transient_1983}.

The initial condition assumes there are no clusters and one mole of liquid monomers. Reflective boundary conditions are employed at $\phi=0,1$, and $n=1$, except for the reaction taking $n=1,\phi=0$ clusters to the fluid. An absorbing boundary is used at $n=100$ to account for the eventual crystallization of large clusters. This only influences the steady state properties of the cluster population down to $n\approx 90$, which is well above $n^*=64$ for the model parameters studied, $\Delta_d=0.4$, $\gamma_c\sigma=6.0$,  and $g=0-0.66$. We choose $C=D=5.0$ so the small cluster behaviour is similar for both models of ordering kinetics and we set $\delta\phi=0.01$.

\section{Lag Times and Steady State}
\begin{figure}[]
\includegraphics[]{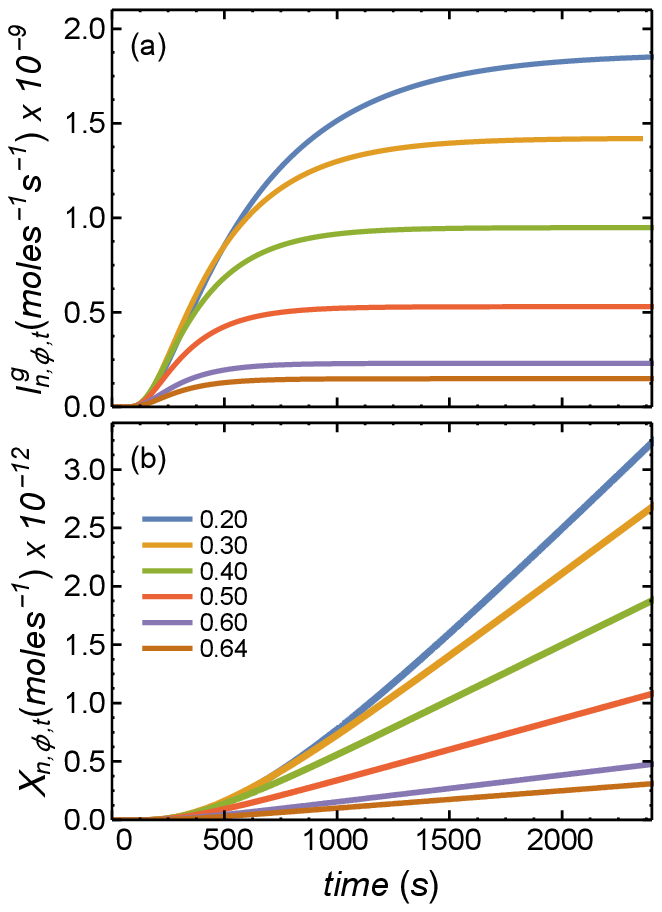}
\caption{(a) $I^g_{n,\phi,t}$, and (b) $\chi_{n,\phi,t}$, as a function of $t$ for critical clusters ($n=64$, $\phi=1.0$), for different values of $g$ in the Surface Mediated Ordering model.}
\label{fig:crates}
\end{figure}
To calculate the lag time, $\tau_{lag}$, we first measure the forward rate for cluster growth, $I^{g}_{n,\phi,t}$ given by Eq.~\ref{eq:fn}, at the critical cluster for the system ($n^*=64$, $\phi=1.0$), then obtain an estimate of the total number of clusters passing through the critical point,
\begin{equation}
\chi_{n,\phi,t^{\prime}}=\int_0^{t^{\prime}}I^{g}_{n,\phi,t} dt\mbox{,}\\
\label{eq:chi}
\end{equation}
where the integral is performed numerically. A linear fit to $\chi_{n,\phi,t}$ is obtained for the data $t>2000$s, and $\tau_{lag}$ obtained as the value of $t$ where the fit extrapolates back to zero. Figure~\ref{fig:crates} shows both $I^{g}_{n,\phi,t}$ and $\chi_{n,\phi,t}$ at the critical cluster as a function of time for different $g$ in the SMO model, which is the slowest model to reach its steady state. For most $g$, $I^{g}_{n,\phi,t}$ exhibits a clear plateau indicating it has indeed reached a stead state at long times which gives rise a linear behaviour in $\chi_{n,\phi,t}$. However, for $g=0.2$ the population is still increasing slowly by the end of the simulation so the linear extrapolation of $\chi_{n,\phi,t}$ leads to an under estimation of the true lag time for this state point.

\section{Cluster Populations}
\begin{figure}[]
\includegraphics[]{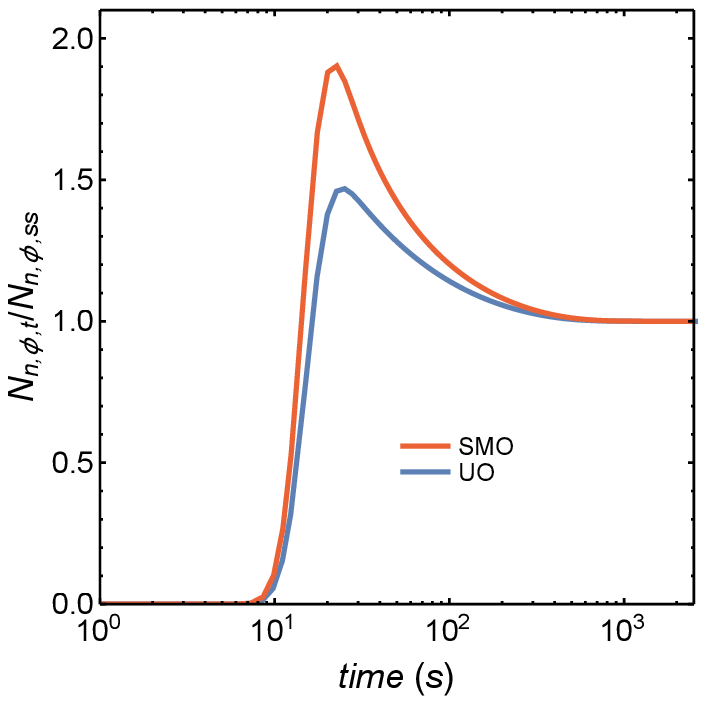}
\caption{Relative transient cluster populations for $n=64$, $\phi=0.02$ clusters as a function of time for the UO and SMO models with $g=0.64$.}
\label{fig:NNsscomp}
\end{figure}

\begin{figure}[t]
\includegraphics[]{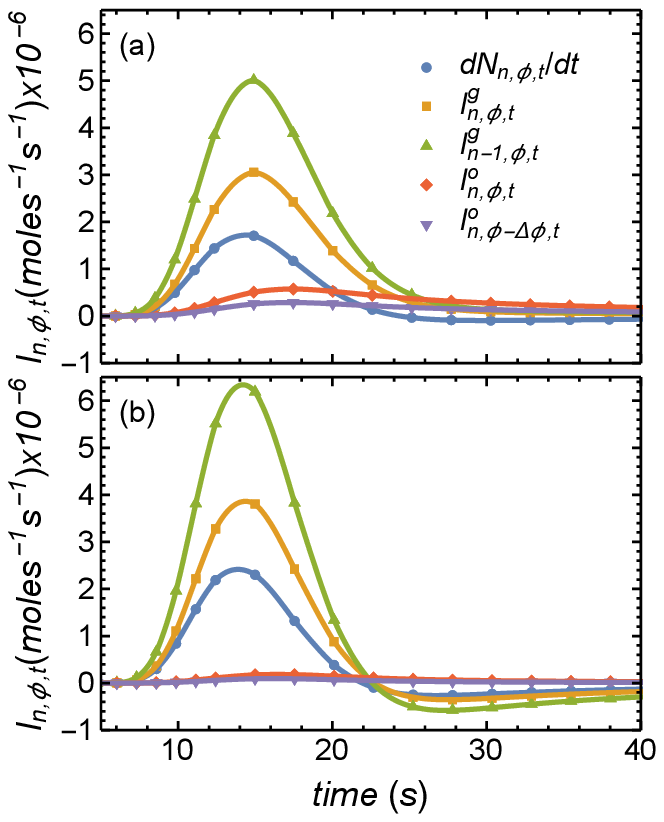}
\caption{$dN_{n,\phi,t}/dt$, and its forward rate contributions (Eq~\ref{eq:dNnt}) at $n=64$ and $\phi=0.02$, for (a) the UO model and (b) the SMO model with $g=0.64$. }
\label{fig:rates}
\end{figure}

A key find of our work is that the cluster population of disordered clusters overshoot their steady state values at intermediates times and then decay to over longer times. Figure~\ref{fig:NNsscomp} shows that this occurs for both models model, which suggests the early accumulation of excess populations of disordered clusters is a general feature of the topology of the free energy surface. The effect is greater for the SMO model because of the slower ordering of the larger particles.  Figures~\ref{fig:rates}(a) and \ref{fig:rates}(b) show $dN_{n,\phi,t}/dt$, and its forward rate contributions given in Eq.~\ref{eq:dNnt}, for large disordered clusters with $n=64$ and $\phi=0.02$, in the UO and SMO models respectively. The peak in $N_{n,\phi,t}$, observed in Fig.~\ref{fig:Nntrans}(b) of the main text, occurs when $dN_{n,\phi,t}/dt=0$, indicating the time at which the cluster excess begins to decline towards its steady state value.  In the UO model, $I^g_{n-1,\phi,t}>I^g_{n,\phi,t}>0$ for all $t$. At early and intermediate times, $I^g_{n-1,\phi,t}$ is the largest rate, leading to the accumulation of disordered clusters but as $I^g_{n-1,\phi,t}$,  $I^g_{n,\phi,t}$ and the difference between the two rates decrease, $I^o_{n,\phi,t}$ becomes the most dominant rate at  the excess cluster population is removed by disordered clusters ordering directly to the crystal state. In the SMO model, the excess cluster population accumulates in the same way. However,  $I^o_{n,\phi,t}$ remains small at all times and we see $I^g_{n-1,\phi,t}$ and $I^g_{n,\phi,t}$ become negative, with the $I^g_{n-1,\phi,t}$ being the largest rate. This suggests the cluster excess dissipates as the disordered clusters begin to shrink in size, while only ordering slowly. The difference in the time evolution of the forward rates ultimately give rise to the distinct nucleation pathways observed in vector fields present in Fig.~\ref{fig:vectors}.

\end{document}